\documentclass[preprintnumbers,aps,prd,reprint,groupedaddress,nofootinbib]{revtex4-1}
\pdfoutput=1

\usepackage{booktabs}
\usepackage{footnote}

\usepackage{lipsum}
\usepackage{amsmath}    
\usepackage{amssymb}    
\usepackage{color}         
\usepackage{graphicx}     
\usepackage[T1]{fontenc}
\usepackage{times}         
\usepackage{mathrsfs} 
\usepackage{myterms}
\usepackage{hyperref} 
\hypersetup{
    colorlinks=true,       
    linkcolor=blue,        
    citecolor=blue,        
    filecolor=magenta,     
    urlcolor=blue          
}
\usepackage[all]{hypcap} 

\usepackage[none]{hyphenat} 
\usepackage{ upgreek } 
\usepackage{booktabs}

\usepackage{color,xcolor,fancybox,epsf,rotating,colordvi}

\usepackage{slashed}
\newcommand{\E}[1]{\times 10^{#1}} %

\begin{document}

\preprint{\parbox{1.6in}{\noindent arXiv:20**.*****}}

\title{Higgsino Asymmetry and Direct-Detection Constraints of Light Dark Matter\\ in the NMSSM with Non-Universal Higgs Masses}

\author{Kun Wang}
\email[]{wk2016@whu.edu.cn}
\affiliation{Center for Theoretical Physics, School of Physics and Technology, Wuhan University, Wuhan 430072, China}

\author{Jingya Zhu}
\email[Corresponding author: ]{zhujy@whu.edu.cn} 
\affiliation{Center for Theoretical Physics, School of Physics and Technology, Wuhan University, Wuhan 430072, China}

\author{Quanlin Jie}
\email[Corresponding author: ]{qljie@whu.edu.cn} 
\affiliation{Center for Theoretical Physics, School of Physics and Technology, Wuhan University, Wuhan 430072, China}

\date{\today}

\begin{abstract}
In this work, we study the direct-detection constraints of light dark matter in the next-to minimal supersymmetric standard model (NMSSM) with non-universal Higgs masses (NUHM), especially the correlation between higgsino asymmetry and spin-dependent cross section.
Finally, we get the following conclusions:
(i) The spin-dependent cross section is proportional to the square of higgsino asymmetry in dark matter $\tilde{\chi}^0_1$ in the NMSSM-NUHM.
(ii) For highly singlino-dominated dark matter $\tilde{\chi}^0_1$, the relic density can be sufficient, but the higgsino asymmetry and spin-dependent cross section is always small.
(iii) With a sizeable higgsino component in the light dark matter, the higgsino asymmetry and spin-dependent cross section can be sizeable, but dark matter relic density is always small, thus it can escape the direct detections.
(iv) Light dark matter in the $h_2$- and $Z$-funnel annihilation channels with sufficient relic density can be covered by future  LUX-ZEPLIN (LZ) 7-ton in spin-dependent detections.
\end{abstract}


\maketitle

\section{Introduction}
\label{sec:introduction}

Dark matter (DM) is always a fascinating topic, as there are so many evidence in astrophysical observations \cite{Bertone:2016nfn, Bertone:2004pz, Jungman:1995df}.
Weakly interacting massive particle (WIMP) is an influential candidate for the cold dark matter, which is generated by the freeze-out production mechanism \cite{Bernstein:1985th, Srednicki:1988ce}.
However, recent results of direct detection of dark matter have set strong constraints to WIMP at the electroweak scale.
To study the correlation between the nature of WIMP DM and its scattering with nucleon, and check the status of possible WIMP dark matter under current direct detection constraints is an urgent task.

Supersymmetry (SUSY) is an attractive concept, which builds an internal symmetry between fermions and bosons and ensures the unification of three gauge interactions.
The $\mu$-problem in the Minimal Supersymmetric extension to the Standard Model (MSSM), which is the simplest implementation of SUSY, can be solved by introducing a complex singlet superfield $\hat{S}$, which is called the Next-to Minimal Supersymmetric Standard Model (NMSSM).
The NMSSM has attracted a lot of attentions especially after the Higgs discovered \cite{Cao:2012fz, Ellwanger:2011aa, King:2012is, Gunion:2012zd, King:2012tr, Ellwanger:2012ke, Gherghetta:2012gb, Gunion:2012gc, Vasquez:2012hn, Agashe:2012zq, Badziak:2013bda, Kowalska:2012gs, Cao:2013gba, King:2014xwa, Cao:2013si, Barbieri:2013hxa, Baglio:2013iia, Christensen:2013dra, Ellwanger:2013ova, Choi:2012he, Nhung:2013lpa, Goodsell:2014pla, Cao:2012yn, Ma:2020mjz}.
The NMSSM with non-universal Higgs masses (NUHM) \cite{Wang:2019biy, Wang:2018vxp, Ellwanger:2014dfa, Das:2013ta, Ellwanger:2018zxt, Ellwanger:2016sur}, which relaxes the Higgs sector at GUT scale and is also known as a semi-constrained NMSSM (scNMSSM),
can satisfy current constraints including Higgs data, LHC searches for sparticles, DM relic density, direct searches for dark matter, and muon g-2, etc \cite{Wang:2020tap}.

In our previous studies of the NMSSM-NUHM \cite{Wang:2020tap}, we developed a novel efficient method to scan the parameter space of NMSSM-NUHM, which consists of the Heuristically Search (HS) and the Generative Adversarial Network (GAN).
We found that NMSSM-NUHM can satisfy all constraints including DM relic density and the new method is efficient.
Then we studied light DM and Higgs invisible decay in the NMSSM-NUHM \cite{Wang:2020dtb}, and found four funnel-annihilation mechanisms for the lightest supersymmetric particle (LSP) $\tilde{\chi}^0_1$, which are the $h_2$, $Z$, $h_1$ and $a_1$ funnel.

In this work, we study in detail the direct detection of light dark matter in the NMSSM-NUHM.
The main difference between spin-independent and spin-dependent scattering is that for the coupling to the nucleon spin which is not simply added coherently by all nucleons inside a nucleus.
The correlation between the spin-independent and spin-dependent direct detection of dark matter has been studied in Ref. \cite{Cohen:2010gj}.
The spin-dependent direct detection of dark matter in MSSM has been studied in Refs. \cite{Bramante:2015una, Catalan:2015cna, Duan:2017ucw, Chakraborti:2017dpu, Profumo:2016zxo}, which is determined by the higgsino sector.
The singlino-higgsino dark matter in the NMSSM has been studied in Refs. \cite{Xiang:2016ndq, Abdallah:2019znp, Badziak:2015exr, Cao:2016nix, Mou:2017sjf, Beskidt:2017xsd, Barman:2020vzm}, and there is another singlet component which is not contained in MSSM.
So in this work, we study especially the correlation between the spin-dependent cross section and the so-called `higgsino asymmetry' in the NMSSM-NUHM.

The rest of this paper is organized as follows.
In \sref{sec:model}, we briefly introduce the model NMSSM-NUHM, especially its electroweakino sector, and some analytic calculations of DM-nucleon scattering.
In \sref{sec:discussion}, we present our numerical results and discussions.
Finally, we draw our conclusions in \sref{sec:conclusion}.

\section{The model and analytic calculations}
\label{sec:model}

The NMSSM extends the MSSM with a singlet superfield $\hat{S}$, in addition to the up- and down-doublet ones $\hat{H}_{u,d}$.
The superpotential with $\mathbb{Z}_3$ symmetry is given as:
\begin{equation}
    W_{\rm NMSSM}=W_{\rm MSSM}^{\slashed{\mu}} + \lambda  \hat{S} \hat{H}_u \cdot \hat{H}_d + \frac{\kappa}{3} \hat{S}^3 \, ,
\end{equation}
where $W_{\rm MSSM}^{\slashed{\mu}}$ is the superpotential of MSSM without the $\mu$-term, and $\lambda$ and $\kappa$ are dimensionless coupling constants.
After electroweak symmetry breaking, the scalar fields $H_{u,d}$ and $S$ get their vacuum expectation values (VEVs) $v_{u,d}$ and $v_s$, with parameters $\tan\beta\equiv v_u/v_d$ and $\mu_{\rm eff}\equiv \lambda v_s$.
Then the gauge eigenstates mix to form 3 CP-even ($h_{1,2,3}$) and 2 CP-odd mass-eigenstate Higgs ($a_{1,2}$).

The soft breaking term in NMSSM is given as:
\begin{eqnarray}
-\mathcal{L}_{\rm NMSSM}^{\rm soft} =
&-&\mathcal{L}_{\rm MSSM}^{\rm soft,\slashed{\mu}} + {M}_S^2 |S|^2 \\ \nonumber
&+&  \lambda A_\lambda S H_u \cdot H_d
+ \frac{\kappa}{3}  A_\kappa S^3 + h.c. \,,
\end{eqnarray}
where ${M}_S^2$ is the soft mass of the singlet field $S$, and $A_{\lambda,\kappa}$ are trilinear couplings with mass dimension.

In the NMSSM-NUHM, the soft masses in the Higgs sector $M^2_{H_{u},H_{d}}$ and $M^2_S$ are assumed to be non-uniform, and they can transform to parameters $\lambda, \kappa, \mu_{\rm eff}$.
Then the model can be determined by following nine parameters
\begin{equation}
    \lambda,\,\, \kappa,\,\, \tan\beta,\,\, \mu_{\rm eff},\,\, A_\lambda,\,\, A_\kappa,\,\, A_0,\,\, M_{1/2}, \,\,M_0  \, ,
\end{equation}
where $M_0$ and $M_{1/2}$ are the uniformed sfermion and gaugino masses, and $A_0$ is the uniformed trilinear coupling in the sfermion sector.

\textbf{The neutralino sector:}

The five super pantners bino $\tilde{B}$, wino $\tilde{W}^{0}$, higgsinos $\tilde{H}_{u,d}$, and singlino $\tilde{S}$ have the same quantum numbers, thus they can mix to five mass-eigenstate neutralinos.
In the base $\psi_i = (\tilde{B}, \tilde{W}^{0}, \tilde{H}_{d}, \tilde{H}_{u}, \tilde{S})$ (with $i=1,2,3,4,5$), the neutralino mass matrix is given by \cite{Ellwanger:2009dp}
\begin{widetext}
\begin{eqnarray}
M_{\tilde{\chi}^{0}}= \left(
\begin{array}{ccccc}
        M_{1}        & 0                 & -\cos \beta  \sin \theta_{\rm W} m_Z  &  \sin \beta  \sin \theta_{\rm W} m_Z  & 0 \\
        0            & M_{2}             &  \cos \beta  \cos \theta_{\rm W} m_Z  & -\sin \beta  \cos \theta_{\rm W} m_Z  & 0 \\
    -\cos \beta  \sin \theta_{\rm W} m_Z &  \cos \beta  \cos \theta_{\rm W} m_Z  & 0               & -\mu_{\rm eff}             & -\lambda v_{d} \\
     \sin \beta  \sin \theta_{\rm W} m_Z & -\sin \beta  \cos \theta_{\rm W} m_Z  & -\mu_{\rm eff}            & 0                & -\lambda v_{u} \\
        0            & 0                 & -\lambda v_{d}      & -\lambda v_{u}   & 2\kappa v_s \,, \\
\end{array}
\right)
\end{eqnarray}
\end{widetext}
where $\theta_W$ is the electroweak mixing angle.

After diagonalization, one can get the mass-eigenstate neutralinos in mass order
\begin{eqnarray}
\tilde{\chi}^0_i = N_{ij} \psi_j \,,
\end{eqnarray}
where $N_{ij}$ means the $\psi_j$ component in $\tilde{\chi}^0_i$, and $\psi_j = (\tilde{B}, \tilde{W}^{0}, \tilde{H}_{d}, \tilde{H}_{u}, \tilde{S})$.

In the previous studies of NMSSM-NUHM, we found that bino and wino are constrained to be very heavy, because of the high mass bounds of gluino, thus they can be decoupled from the light sector, unlike that in the non-universal gaugino masses scenario \cite{Wang:2018vrr}.
Then the following approximations for $N_{ij}$ can be found \cite{Cao:2015loa, Badziak:2015exr}:
\begin{eqnarray}
&&N_{i3}:N_{i4}:N_{i5}= \\ \nonumber
&&\left[ \frac{m_{\tilde{\chi}^0_{ i}}}{\mu_{\rm eff}} \sin \beta - \cos \beta\right]
: \left[ \frac{m_{\tilde{\chi}^0_{ i}}}{\mu_{\rm eff}} \cos \beta - \sin \beta\right]
: \frac{\mu_{\rm eff} - m_{\tilde{\chi}^0_{ i}}}{\lambda v}
\end{eqnarray}
When we neglect the bino and wino components in the LSP $\tilde{\chi}^0_1$, the singlino and higgsino components of LSP might have the relation
\begin{equation}
    \label{eq:=1}
    N_{13}^2+N_{14}^2+N_{15}^2 \approx 1 \, .
\end{equation}
Since $\tan \beta \gg 1$, we can have $\sin \beta \approx 1$ and $\cos \beta \approx 0$, then for components in LSP we can write
\begin{eqnarray}
    \label{eq:1314}
N_{13}:N_{14}:N_{15}=
\frac{m_{\tilde{\chi}^0_{ 1}}}{\mu_{\rm eff}}
:   - 1
: \frac{\mu_{\rm eff} - m_{\tilde{\chi}^0_{ 1}}}{\lambda v} \, .
\end{eqnarray}


Since $\tilde{B}, \tilde{W}^{0}, \tilde{S}$ are SU(2) singlet, they have no interaction with $Z$ boson, then the coupling between $Z$ and $\tilde{\chi}^0_{\rm 1}$ is given as
\begin{eqnarray}
C_{Z \tilde{\chi}_1^{0}  \tilde{\chi}_1^{0} } =
&&-\frac{i}{2} (g_1 \sin\theta_W \!+\! g_2 \cos\theta_W)(N^2_{1 3} \!-\! N^2_{1 4} )\gamma_{\mu} P_L      \nonumber\\
&& + \frac{i}{2} (g_1 \sin\theta_W \!+\! g_2 \cos\theta_W)(N^2_{1 3}\!-\! N^2_{1 4})\gamma_{\mu} P_R \,,~~~~~~
\label{eq:higgsino}
\end{eqnarray}
which means the coupling is proportional to the so-called `higgsino asymmetry' $|N^2_{1 3}- N^2_{1 4}|$ \cite{Caron:2015wda, Profumo:2017hqp}.

\begin{figure}[htbp]
    \includegraphics[width=.38\textwidth]{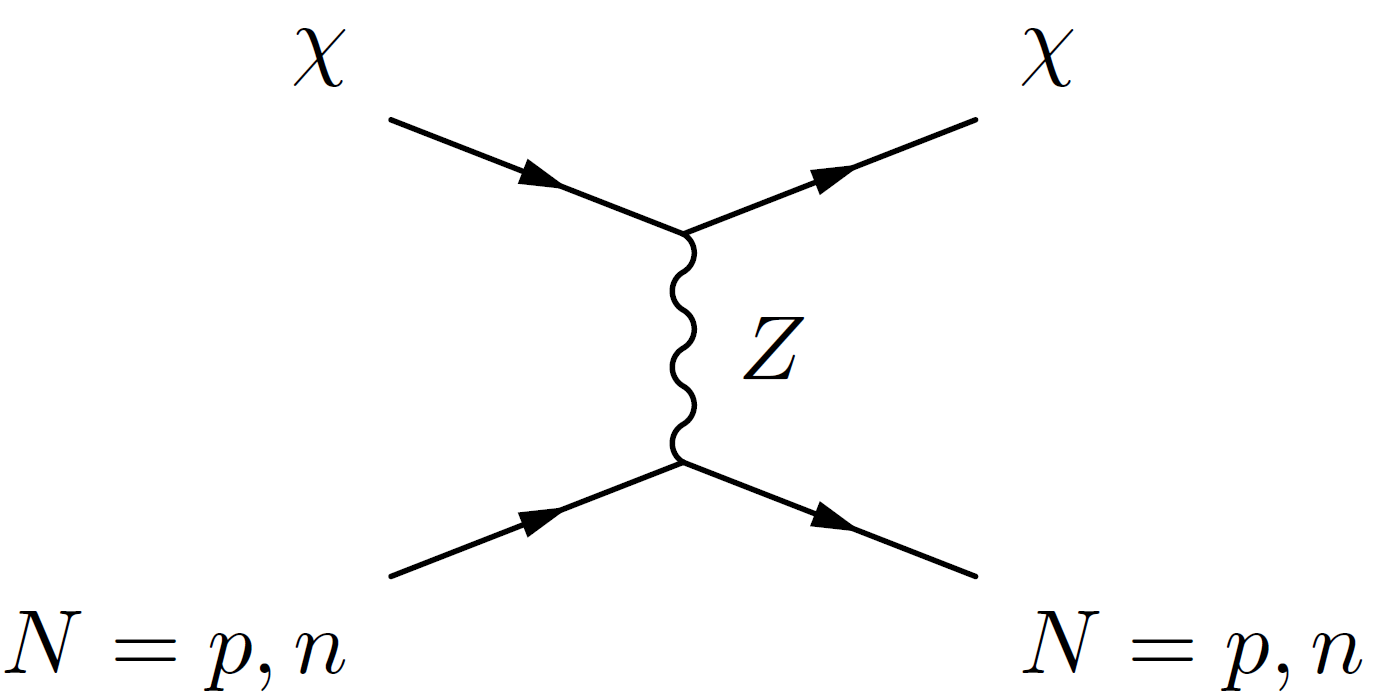}
    \caption{\label{fig:z}
    The scattering of dark matter and nucleons, mediated by $Z$ boson, where the $\chi$ represents dark matter and the $N$ represents nucleons, including protons $p$ and neutrons $n$.
    }
\end{figure}

For WIMP dark matter, such as LSP $\tilde{\chi}^0_1$ in the NMSSM, the principle of direct detections is using its scattering with nucleons.
The $\tilde{\chi}^0_1$-nucleon scattering cross section can be described by a effective four-fermion operator $g_{N N \tilde{\chi}_{1}^{0} \tilde{\chi}_{1}^{0}}$, just like that in Fermi's theory.
For spin-independent (SI) scattering, the dominant contribution is from the exchange of a Higgs boson, since squarks are heavy.
While for spin-dependent (SD) scattering, the dominant contribution is from the exchange of the $Z$ boson, as shown in Fig.\ref{fig:z}.

For LSP $\tilde{\chi}_{1}^{0}$ in the NMSSM-NUHM, the SD cross section is given as \cite{Plehn:2017fdg}
\begin{eqnarray}
    \sigma_{\mathrm{SD}}\left(\tilde{\chi}_{1}^{0} N \rightarrow \tilde{\chi}_{1}^{0} N\right) \approx \frac{4 g_{N N \tilde{\chi}_{1} \tilde{\chi}_{1}^{0}}^{2} }{\pi} \frac{m_{N}^{2}}{\Lambda^{4}}
\end{eqnarray}
Combining with the Eq.(\ref{eq:higgsino}), and considering the SM coupings are constant, one can get the correlation between the SD cross section and the higgsino asymmetry
\begin{eqnarray}
    \label{eq:square}
    \sigma_{\mathrm{SD}}\left(\tilde{\chi}_{1}^{0} N \rightarrow \tilde{\chi}_{1}^{0} N\right) \propto |N^2_{1 3}- N^2_{1 4}|^2
\end{eqnarray}

The detail calculation of dark matter observables in the NMSSM is calculated with \textsf{micrOMEGAs 5.0} \cite{Belanger:2006is,Belanger:2008sj,Belanger:2010pz,Belanger:2013oya}, including DM relic density, DM direct detection, and annihilation processes.

\section{Numerical calculations and discussions}
\label{sec:discussion}

\begin{figure*}[!htbp]
    \includegraphics[width=1.0\textwidth]{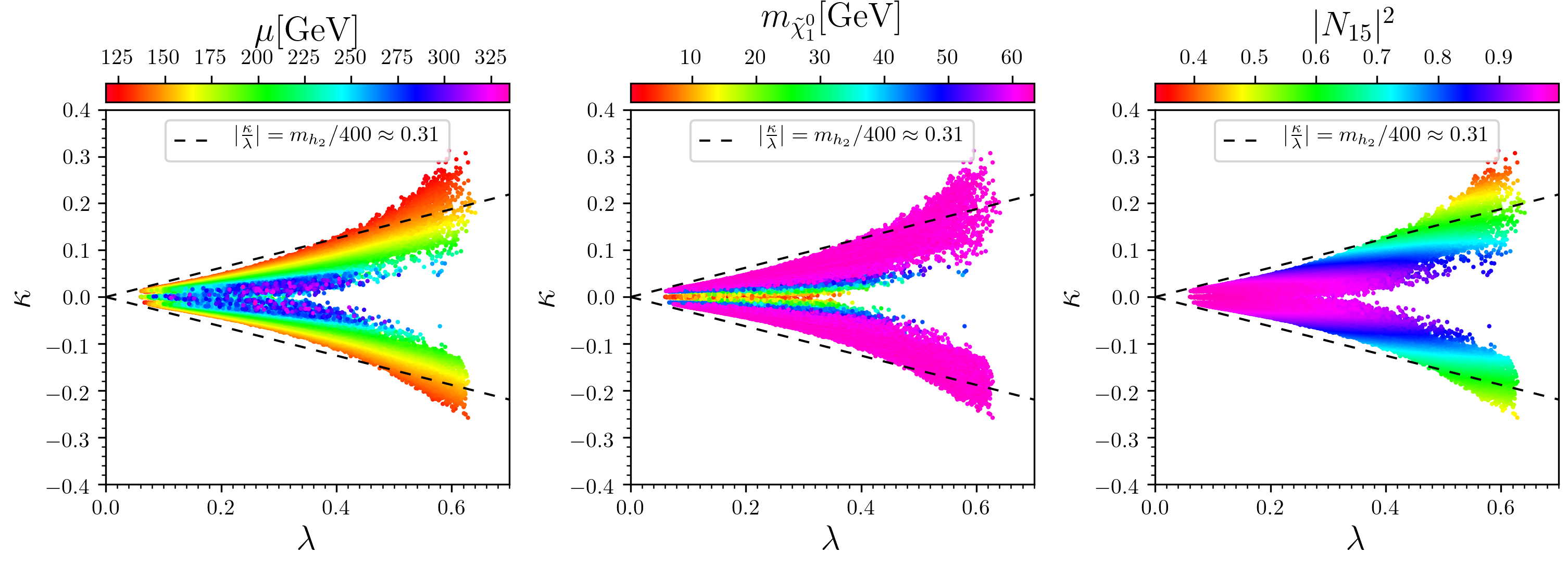}
    \vspace{-0.8cm}
    \caption{\label{fig:1}
    The surviving samples in the $\kappa$ versus $\lambda$ planes.
    From left to right, colors indicate $\mu$, mass of the LSP  $m_{\tilde{\chi}^0_1}$, and the singlino component in the LSP $|N_{15}|^2$, respectively.
    The dashed line $|\kappa/\lambda|=125/400\approx0.31$.
    In all these samples, samples with larger $\mu$ are projected on the top of smaller ones.
    }
\end{figure*}

\begin{figure*}[!htbp]
    \includegraphics[width=1.0\textwidth]{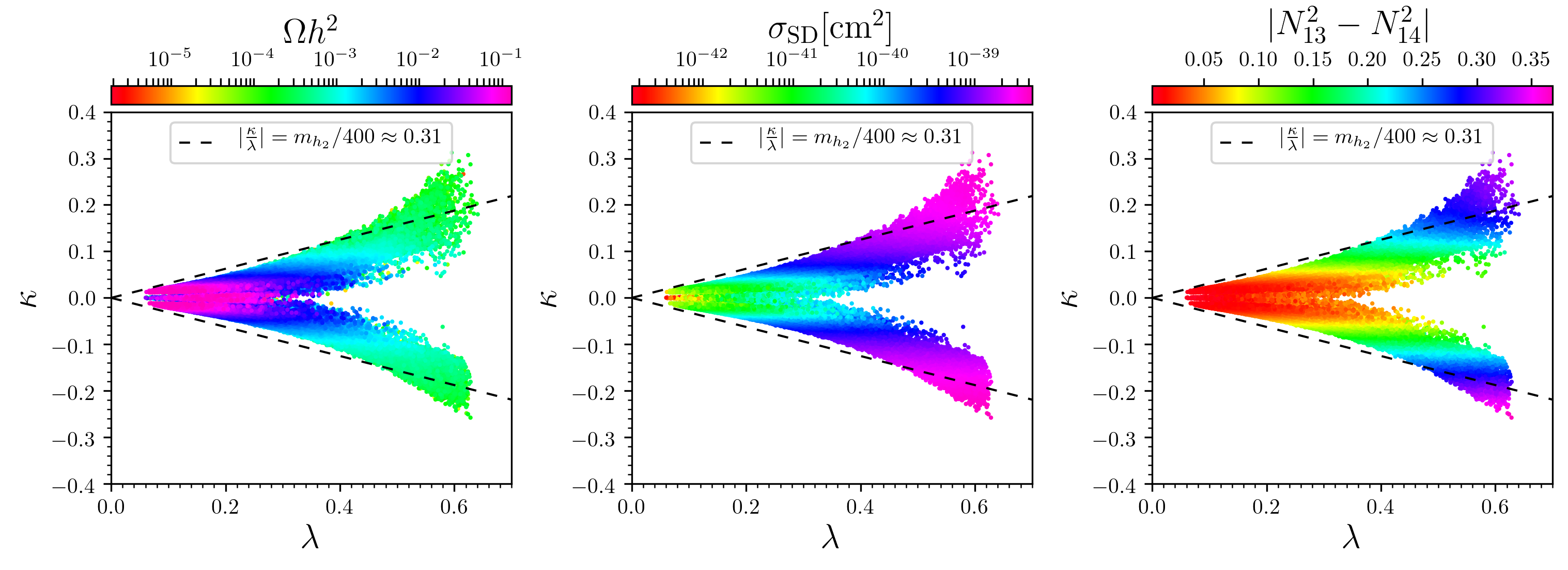}
    \vspace{-0.8cm}
    \caption{\label{fig:2}
    The surviving samples in the $\kappa$ versus $\lambda$ planes.
    From left to right, colors indicate the DM relic density $\Omega h^2$, spin-dependent DM-neutron cross section $\sigma_{\rm SD}$, and the Higgsino asymmetry $| N_{13}^2-N_{14}^2|$, respectively.
    The dashed line is the same as in Fig.\ref{fig:1}.
    In all these samples, samples with larger DM relic density $\Omega h^2$ is projected on the top of smaller ones.
    }
\end{figure*}

In this work, we study in detail the direct-detection constraints to light dark matter $\tilde{\chi}^0_{\rm 1}$, whose mass is lighter than half of the SM-like Higgs $m_{h_2}$.
Since higgsino asymmetry correlates with the SD detection of dark matter, we extend the region of $\mu_{\rm eff}$ to 100-1000 GeV and updated our scan result in Ref. \cite{Wang:2020tap}.
In the scan we use \textsf{NMSSMTools-5.5.2} \cite{Ellwanger:2004xm, Ellwanger:2005dv, Ellwanger:2006rn} to generate the particle spectrum, imposing the corresponding constraints there, use \textsf{HiggsBounds-5.8.0} \cite{Bechtle:2015pma,Bechtle:2013wla,Bechtle:2013gu,Bechtle:2011sb,Bechtle:2008jh} to constrain the extra Higgs bosons, and use \textsf{SModelS-v1.2.4} \cite{Kraml:2013mwa,Ambrogi:2017neo,Ambrogi:2018ujg,Dutta:2018ioj,Buckley:2013jua,Sjostrand:2006za,Sjostrand:2014zea,Alguero:2020grj} to constrain the sparticle sectors.
Thus eventually the surviving samples shown hereafter satisfy the constraints including Higgs data, muon g-2, B physics, sparticle searches, etc.
Specifically, the detail of dark matter constraints are list as following:
\begin{itemize}
    \item The upper bound of DM relic density $\Omega h^2 \le 0.131$ \cite{Tanabashi:2018oca,Hinshaw:2012aka,Ade:2013zuv};
    \item The SI DM-nucleon cross section is constrained by XENON1T \cite{Aprile:2018dbl}, which is rescaled by $\Omega/\Omega_0$ with $\Omega_0 h^2=0.1187$;
    \item The SD DM-proton cross section is constrained by LUX \cite{Akerib:2016lao}, XENON1T \cite{Aprile:2019dbj} and PICO-60 \cite{Amole:2019fdf}, which is also rescaled by $\Omega/\Omega_0$.
\end{itemize}

Similar to that described in Ref.\cite{Wang:2020tap}, the mass regions of most particles in this scenario is shown in Tab.\ref{tab:spectrum}.
In the following, we focus on the light LSP $\tilde{\chi}^0_{\rm 1}$ in the NMSSM-NUHM: its scattering cross sections with nucleons and their correlations with higgsino asymmetry.

\begin{table}[!htb]
\caption{\label{tab:spectrum}
The mass region of most particles and some related parameters.
        }
\begin{ruledtabular}
\begin{tabular}{@{}ccc@{}}
\toprule
mass/parameter & minimum & maximum \\ \midrule
$M_{1/2} ~~[\GeV]$             & 1045 & 1564 \\
$M_{\tilde{g}} ~~[\GeV]$            & 2296 & 3328 \\
$M_{\tilde{q}_{1,2}} ~~[\GeV]$            & 2000 & 3021 \\ \hline
$M_2 ~(M_{\rm SUSY}) ~~[\GeV]$          & 825  & 1238 \\
$M_1 ~(M_{\rm SUSY}) ~~[\GeV]$          & 451  & 683  \\
$\mu_{\rm eff} ~~[\GeV]$          & 118  & 335  \\
$| 2 \kappa \cdot \mu_{\rm eff} / \lambda | ~~[\GeV]$   & 0.3  & 62   \\ \hline
$m_{\tilde{\chi}^0_1} ~~[\GeV]$          & 0.4  & 63   \\
$m_{\tilde{\chi}^0_2} ~~[\GeV]$          & 123 & 331  \\
$m_{\tilde{\chi}^0_3} ~~[\GeV]$          & 133 & 348  \\
$m_{\tilde{\chi}^\pm_1} ~~[\GeV]$          & 120  & 337  \\
$m_{\tilde{\chi}^0_4} ~~[\GeV]$          & 449  & 675  \\
$m_{\tilde{\chi}^0_5}, ~m_{\tilde{\chi}^\pm_2} ~~[\GeV]$          & 852  & 1271 \\ \hline
$m_{\tilde{\uptau}_1} ~~[\GeV]$           & 93   & 608  \\
$m_{\tilde{\nu}_\uptau} ~~[\GeV]$           & 62   & 644  \\
$m_{\tilde{\mu}_1} ~~[\GeV]$  & 382  & 830  \\
$m_{\tilde{\nu}_\mu} ~~[\GeV]$            & 375  & 826  \\
$m_{\tilde{t}_1} ~~[\GeV]$           & 700  & 2164 \\
$m_{h_1} ~~[\GeV]$         & 2.5 & 111  \\
$m_{a_1} ~~[\GeV]$         & 0.1 & 191  \\ \bottomrule
\end{tabular}
\end{ruledtabular}
\end{table}

\begin{figure*}[!htbp]
    \includegraphics[width=.66\textwidth]{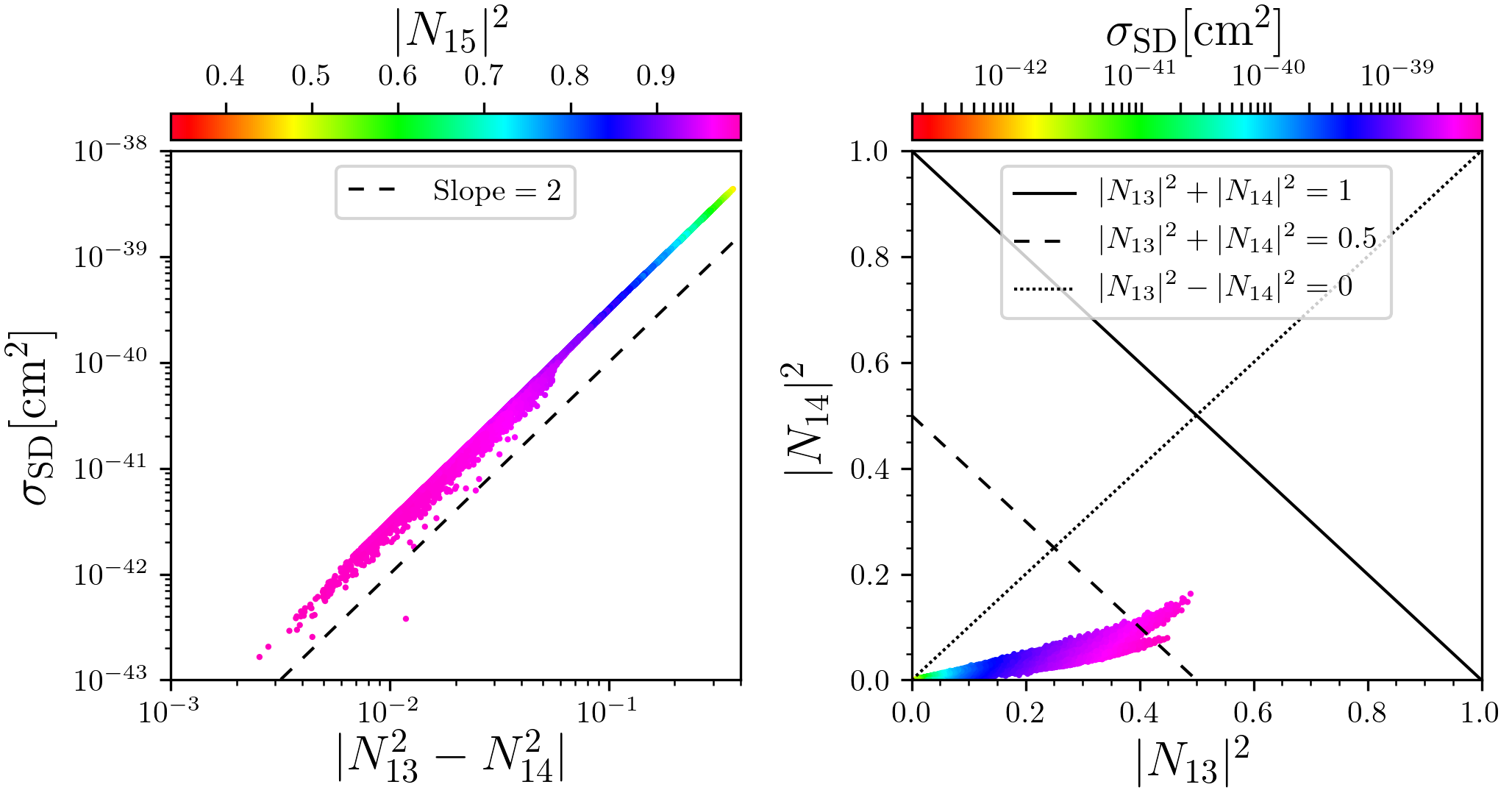}
    \vspace{-0.5cm}
    \caption{\label{fig:3}
    Surviving samples in the spin-dependent DM-neutron cross section $\sigma_{\rm SD}$ versus the higgsino asymmetry $|N_{13}^2-N_{14}^2|$ (left), and higgsino components in LSP $|N_{13}|^2$ versus $|N_{14}|^2$ planes, with colors indicating the singlino component in the LSP $|N_{15}|^2$ and spin-dependent WIMP-neutron cross section $\sigma_{\rm SD}$, respectively.
    In the left panel, the slope of the black dashed line is 2.
    In the right panel, the black solid, dashed and dotted lines indicate the $|N_{13}|^2+|N_{14}|^2=1$, $|N_{13}|^2+|N_{14}|^2=0.5$ and $|N_{13}|^2=|N_{14}|^2$, respectively.
    }
\end{figure*}

\begin{figure*}[!htbp]
    \includegraphics[width=.66\textwidth]{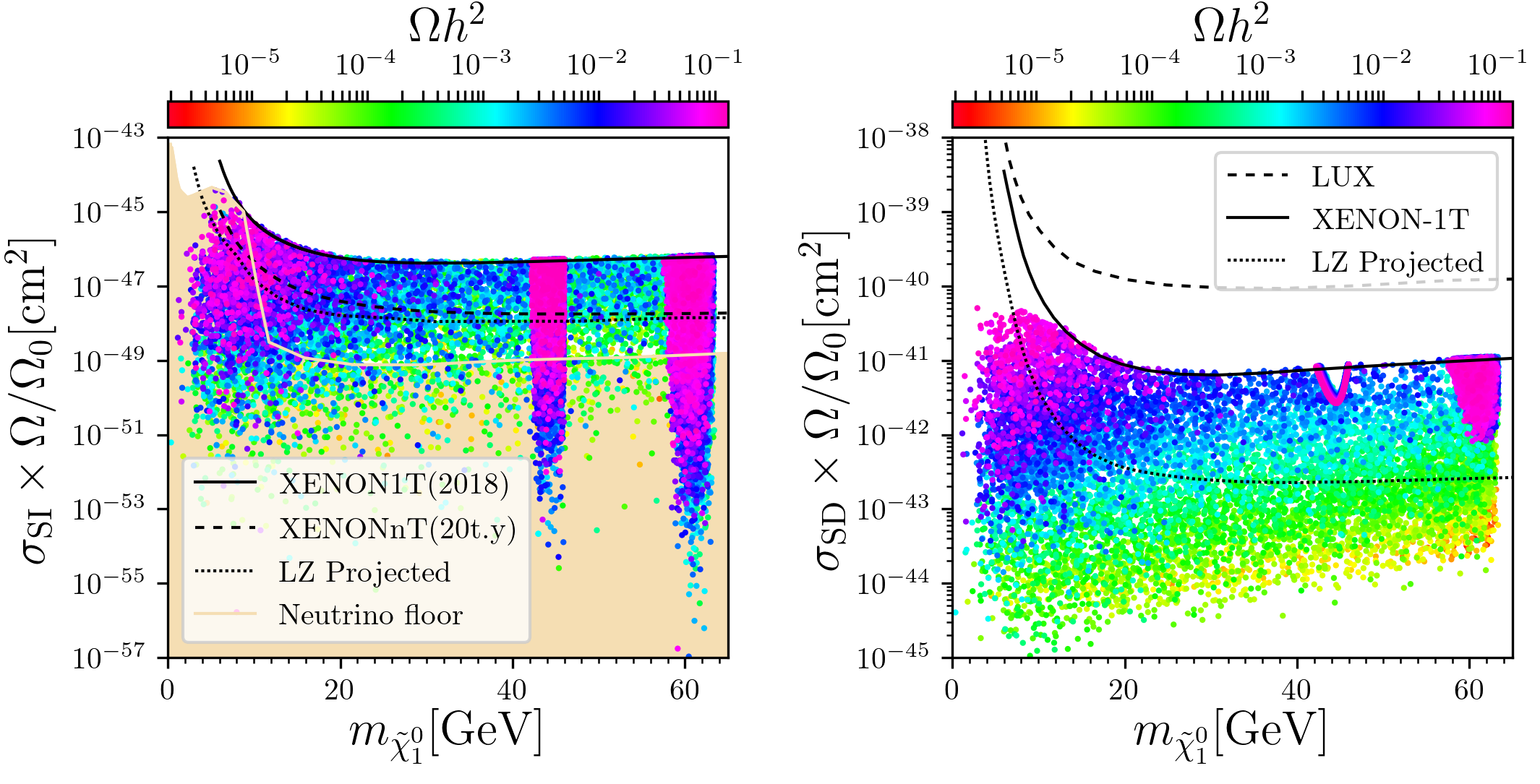}
    \vspace{-0.5cm}
    \caption{\label{fig:4}
    Surviving samples in the rescaled spin-independent $\sigma_{\rm SI} \times \Omega/\Omega_0$ (left) and spin-dependent cross section $\sigma_{\rm SD} \times \Omega/\Omega_0$ (right), versus mass of the LSP $m_{\tilde{\chi}^0_1}$ planes respectively, with colors indicating DM relic density $\Omega h^2$.
    In the left panel, the black solid, dashed, and dotted curves indicate the spin-independent detection limit by XENON1T (2018) \cite{Aprile:2018dbl}, future experiments of XENONnT and LUX-ZEPLIN (LZ), respectively.
    And the neutrino floor \cite{Billard:2013qya} is shown by the orange shaded region.
    In the right panel, the dashed, solid, and dotted curves indicate the spin-dependent detection limit by LUX \cite{Akerib:2016lao}, XENON1T (2018) \cite{Aprile:2019dbj}, and future LZ projected \cite{Akerib:2018lyp} respectively.
    Samples with larger DM relic density $\Omega h^2$ are projected on top of smaller ones.
    }
\end{figure*}

In Fig.\ref{fig:1}, the surviving samples were projected in the $\kappa$ versus $\lambda$ planes, with colors indicating $\mu_{\rm eff}$, mass of the LSP $m_{\tilde{\chi}^0_1}$, and the singlino component $|N_{15}|^2$ respectively.
The dashed lines indicate $\kappa/\lambda=125/400\approx0.31$, which means for the inner samples the singlino-like neutralino is lighter than half of SM-like Higgs mass, since the minimum of $\mu_{\rm eff}$ is about $100\GeV$.
From this figure, we can have the following observations:
\begin{itemize}
    \item The maximum of the $\mu_{\rm eff}$ is about $335 \GeV$. combining with Tab.\ref{tab:spectrum}, we can see this also approximates the maximum of $\tilde{\chi}^0_{2,3}$ and $\tilde{\chi}^\pm_1$, since they are mainly higgsino-dominated.
    \item For samples with large $\mu_{\rm eff}$, $\tilde{\chi}^0_1$ are highly singlino-dominated with mass $m_{\tilde{\chi}^0_1}\approx 2|\kappa\mu_{\rm eff}/\lambda|$, since they are with very small $|\kappa/\lambda|$.
    \item Most of the surviving samples are between the two dashed lines, which means these samples are singlino-dominated.
\end{itemize}

\begin{table*}[!htb]
\caption{\label{tab:benchmark}
Related information for the the six benchmark points in the NMSSM-NUHM, where $|N_{13}|^2$, $|N_{14}|^2$, and $|N_{15}|^2$ are the up-type higgsino, down-type higgsino, and singlino components of the LSP $\tilde{\chi}^0_1$.
        }
\begin{ruledtabular}
\begin{tabular}{@{}ccccccc@{}}
\toprule
            & P1        & P2        & P3        & P4        & P5       & P6        \\  \hline 
$\lambda$    & $9.04\E{-2}$ & $2.81\E{-1}$  & $2.68\E{-1}$  & $2.59\E{-1}$  & $5.95\E{-1}$ & $6.28\E{-1}$  \\
$\kappa $    & -$1.45\E{-2}$& -$6.27\E{-3}$ & -$2.33\E{-3}$ & -$1.36\E{-2}$ & $3.12\E{-1}$ & -$2.58\E{-1}$ \\
$\mu_{\rm eff} \;\rm [GeV]$    & 140       & 285       & 195       & 335       & 118      & 134       \\
$m_{\tilde{\chi}^0_1 } \;\rm [GeV]$     & 46        & 12        & 3         & 34        & 61       & 62        \\
$|N_{13}|^2$     & 1.4\%     & 2.7\%     & 5.1\%     & 1.8\%     & 49\%     & 45\%      \\
$|N_{14}|^2$     & 0.1\%     & 0.01\%    & 0.0008\%  & 0.04\%    & 16\%     & 8\%       \\
$|N_{15}|^2$     & 98.4\%    & 97.2\%    & 94.8\%    & 98.2\%    & 34\%     & 47\%      \\
$|N_{13}^2-N_{14}^2|$ & 1.3\%     & 2.7\%     & 5.1\%     & 1.7\%     & 33\%     & 37\%      \\
$\Omega h^2$       & 0.1114    & 0.1290    & 0.1220    & $1.37\E{-3}$  & $2.52\E{-4}$ & $1.84\E{-4}$  \\
$\sigma_{\rm SI} \;\rm [cm^2]$     & $1.49\E{-49}$  & $1.17\E{-47}$  & $5.62\E{-49}$  & $2.78\E{-46}$  & $2.13\E{-44}$ & $1.22\E{-44}$  \\
$\sigma_{\rm SD} \;\rm [cm^2]$     & $5.45\E{-42}$  & $2.11\E{-41}$  & $4.92\E{-41}$  & $9.23\E{-42}$  & $3.39\E{-39}$ & $4.34\E{-39}$  \\ \bottomrule
\end{tabular}
\end{ruledtabular}
\end{table*}

In Fig.\ref{fig:2}, we show the surviving samples in the $\kappa$ versus $\lambda$ planes, with colors indicate the DM relic density $\Omega h^2$, SD DM-neutron cross section $\sigma_{\rm SD}$, and the higgsino asymmetry $|N_{13}^2-N_{14}^2|$ respectively.
We have checked that the SD DM-proton cross section is almost the same as the SD DM-neutron cross section.
Six benchmark points are listed in Table.\ref{tab:benchmark}.
From this figure, we can have the following observations:
\begin{itemize}
    \item The samples with enough DM relic density is highly singlino-dominated, with smaller $\kappa$ and $\lambda$, with smaller spin-dependent DM-neutron cross section $\sigma_{\rm SD}$ and with smaller higgsino asymmetry.
    \item From the middle and right planes in Fig.\ref{fig:2}, we can see that the SD DM-neutron cross section $\sigma_{\rm SD}$ is proportional to the higgsino asymmetry. And we also checked that the spin-independent DM-nucleon cross section $\sigma_{\rm SI}$ have no such correlation.
    \item For samples with enough DM relic density, since $\tilde{\chi}^0_1$ are with very tiny higgsino components, the asymmetry between higgsino components, $|N_{13}^2-N_{14}^2|$, can not be large, thus the SD DM-neutron cross section $\sigma_{\rm SD}$ is also very tiny.
\end{itemize}

In the Fig.\ref{fig:3}, we show the detail correlation between SD DM-neutron cross section $\sigma_{\rm SD}$ and higgsino asymmetry $|N_{13}^2-N_{14}^2|$.
From this figure, we can have the following observations:
\begin{itemize}
    \item In the left log-log panel, the slope of the black dashed line is 2.
    Since the surviving samples distribute parallel to the line, the correlation between the SD DM-neutron cross section and the higgsino asymmetry is:
    \begin{eqnarray}
        \sigma_{\mathrm{SD}} \propto |N^2_{1 3}- N^2_{1 4}|^2 \, .
    \end{eqnarray}
    \item From the left panel, we can also see that the higgsino asymmetry is anti-correlated with the singlino component in $|N_{15}|^2$.
        Combining with Fig.\ref{fig:1} and \ref{fig:2}, we know the SD cross section is anti-correlated with the relic density, thus the samples can escape the constraints of SD direct-detection even if the higgsino asymmetry is large.
    \item From the right panel, we can see that for most samples $|N_{13}|^2 \ll |N_{14}|^2\lesssim 0.5$.
    \item  Since $118\lesssim\mu_{\rm eff}\lesssim335\GeV$ and $m_{\tilde{\chi}^0_1}\lesssim 63\GeV$ according to Tab.\ref{tab:spectrum}, according to Eq.\eqref{eq:1314}, the component ratio in the LSP $\tilde{\chi}^0_1$ can be approximate to
    \begin{eqnarray}
        \label{eq:higgsino ratio}
    N_{13}:N_{14} \approx
    \frac{m_{\tilde{\chi}^0_{ 1}}}{\mu_{\rm eff}}
    :   - 1  \, ,
    \end{eqnarray}
    so the higgsino asymmetry can not be too small unless the higgsino components are small.
    But when higgsino components are large, the singlino component is small and the relic density is not sufficient.
\end{itemize}

In Fig.\ref{fig:4}, we show the DM direct-detection constraints.
The SI DM-nucleon cross section $\sigma_{\rm SI}$ and SD DM-neutron cross section $\sigma_{\rm SD}$ are both rescaled by a ratio $\Omega/\Omega_0$ with $\Omega_0 h^2=0.1187$.
We can have the following observations:
\begin{itemize}
    \item The constraints from the SI DM-nucleon cross section is much more stringent than those from the SD DM-neutron cross section.
    The reason is that the SI DM-nucleon cross section $\sigma_{\rm SI}$ is is enhanced by the square of the nucleon number in the nucleus, and the SD DM-neutron cross section $\sigma_{\rm SD}$ have no such enhancement.
    \item The SD cross section of the surviving samples is much larger than the SI cross section.
    This is because the LSP $\tilde{\chi}^0_{ 1}$ in NMSSM-NUHM is a Majorana fermion.
    The SD scattering can have a Z boson exchange channel, while the SI scattering is mainly through Higgs bosons exchange.
    As seen from the left and right panels, the SD cross section $\sigma_{\rm SD}$ is larger than SI cross section $\sigma_{\rm SI}$ by about 3-6 orders.
    \item From Fig.\ref{fig:1} and Fig.\ref{fig:2}, we can see that for samples with large SD DM-neutron cross section, the LSP $\tilde{\chi}^0_{ 1}$ are higgsino-dominated and with large higgsino asymmetry.
        From the right panel of Fig.\ref{fig:4}, we can see that due to very small DM relic density, about $10^{-3}-10^{-5}$, they have small rescaled cross sections and can easily escape the direct-detection constraints.
    \item From the right panel, we can see that surviving samples in the $h_2$- and $Z$-funnel annihilation channels with right relic density can be covered by future LZ 7-ton in spin-dependent detections.
\end{itemize}


\section{Conclusions}
\label{sec:conclusion}

In this work, we studied the direct-detection constraints of light dark matter in the NMSSM with non-universal Higgs masses (NMSSM-NUHM), especially the correlation between higgsino asymmetry and spin-dependent cross section.
We first updated the scan result in our previous work \cite{Wang:2020tap} with broader $\mu_{\rm eff}$ regions.
We consider the related theoretical and experimental constraints including Higgs data, muon g-2, B physics, sparticle searches at the LHC, DM relic density, etc., and pay special attention to that of direct detections of light DM.

In the NMSSM-NUHM scenario we consider in this work, the bino and wino are decoupled from the sector of light neutralinos, and the lightest neutralino $\tilde{\chi}^0_1$, as the dark matter lighter than half of the SM-like Higgs, is mixed by singlino and higgsinos.
We studied properties of the dark matter $\tilde{\chi}^0_1$, especially its constraints of direct detection.
Finally, we get the following conclusions:
\begin{itemize}
	\item The spin-dependent cross section is also proportional to the square of higgsino asymmetry in dark matter $\tilde{\chi}^0_1$ in the NMSSM-NUHM.
	\item For highly singlino-dominated LSP, the relic density can be sufficient, but the higgsino asymmetry and spin-dependent cross section is always small.
	\item With a sizeable higgsino component in the light dark matter, the higgsino asymmetry and spin-dependent cross section can be sizeable, but dark matter relic density is always small, thus it can escape the direct detections.
	\item Light dark matter in the $h_2$- and $Z$-funnel annihilation channels with sufficient relic density can be covered by future LZ 7-ton in spin-dependent detections.
\end{itemize}

\begin{acknowledgments}
	\paragraph*{Acknowledgements.}
	This work was supported by the National Natural Science Foundation of China (NNSFC)
	under grant Nos. 11605123.
\end{acknowledgments}

\bibliography{ref}

\end{document}